\documentclass{article}

\usepackage{arxiv}

\usepackage[utf8]{inputenc}
\usepackage[T1]{fontenc}
\usepackage{hyperref}
\usepackage{url}
\usepackage{booktabs}
\usepackage{amsfonts}
\usepackage{amsmath}
\usepackage{amssymb}
\usepackage{nicefrac}
\usepackage{microtype}
\usepackage{cleveref}
\usepackage{graphicx}
\usepackage[numbers,sort&compress]{natbib}
\usepackage{doi}
\usepackage{bm}
\usepackage{caption}
\usepackage{multirow}

\newcommand{\R}{\mathbb{R}}
\newcommand{\vu}{\bm{u}}
\newcommand{\vv}{\bm{v}}
\newcommand{\vw}{\bm{w}}
\newcommand{\vd}{\bm{d}}
\newcommand{\vs}{\bm{s}}
\newcommand{\vq}{\bm{q}}
\newcommand{\vf}{\bm{f}}
\newcommand{\vr}{\bm{r}}
\newcommand{\vphi}{\bm{\phi}}
\newcommand{\vtheta}{\bm{\theta}}
\newcommand{\veps}{\bm{\varepsilon}}
\newcommand{\vsig}{\bm{\sigma}}
\newcommand{\mB}{\bm{B}}
\newcommand{\mD}{\bm{D}}
\newcommand{\mC}{\bm{C}}
\newcommand{\mK}{\bm{K}}
\newcommand{\mM}{\bm{M}}
\newcommand{\mI}{\bm{I}}
\newcommand{\mT}{\bm{T}}
\newcommand{\mS}{\bm{S}}
\newcommand{\mRot}{\bm{R}}
\newcommand{\mRres}{\bm{R}_{\mathrm{res}}}
\newcommand{\mSp}{\bm{S}'}

\title{Joint Nonlinearity in a Stiffened Aluminium Wingbox Panel and What It Requires of a Reduced Basis}

\author{
  Nikolaos D.~Tantaroudas\thanks{Corresponding author, e-mail \texttt{nikolaos.tantaroudas@iccs.gr}} \\
  Institute of Communication and Computer Systems (ICCS) \\
  9, Iroon Politechniou str., Polytechnic Campus \\
  15773 Zografou, Athens, Greece
  \and
  Evangelos Papatheou \\
  Department of Engineering \\
  University of Exeter \\
  Exeter, EX4 4QF, United Kingdom \\
  \texttt{e.papatheou@exeter.ac.uk}
}

\renewcommand{\shorttitle}{Joint nonlinearity and reduction of a wingbox panel}

\hypersetup{
  pdftitle  = {Joint Nonlinearity in a Stiffened Aluminium Wingbox Panel and What It Requires of a Reduced Basis},
  pdfauthor = {Nikolaos D. Tantaroudas, Evangelos Papatheou},
  pdfkeywords = {wingbox panel, shell finite element, joint modelling, nonlinear dynamics, harmonic balance, model order reduction, residual flexibility}
}

\begin{document}
\maketitle

\begin{abstract}
A shell finite-element model of a stiffened aluminium wingbox panel, a laboratory structure shared by several experimental studies, is built with its plate-to-stiffener joints represented at the $78$ physical fastener positions, calibrated against measured modal data, and then used to ask what joint nonlinearity does to the panel and what it demands of a reduced-order model.
The stiffness and mass are assembled from four-node mixed-interpolation flat-shell elements, the fasteners are rigid offset links or compliant elements at the drawing positions, and a six-parameter sensitivity update of the substructure moduli within $\pm20\%$ reproduces the first five measured modes to $1.1$--$4.3\%$ with modal-assurance values of $0.88$--$1.00$.
Making the fasteners nonlinear, at equal mesh, mass and damping, shows that joint nonlinearity enters the response strongly asymmetrically. Hardening the fasteners is almost invisible, at most $+1\%$ in frequency, whereas softening or slipping them moves the first three resonances by $-2.1$, $-4.9$ and $-9.2\%$, and friction removes up to $80\%$ of the resonant peak at intermediate levels before it recovers. Two independent solutions, Newmark integration of all $18{,}804$ degrees of freedom and a harmonic balance condensed exactly onto the joint degrees of freedom and continued in arclength, agree to $5.5\times10^{-4}$.
A projection-based nonlinear model order reduction of the same model then locates the criterion that a jointed structure imposes on a reduced basis. The binding quantity is not the linear response, which eleven vectors reproduce to better than $0.01\%$, nor the joint deformation the shaker produces, but the receptance of the structure between the joints, of which $126$ global eigenvectors carry about $2\%$, and without which the reduced model overpredicts the hardening shift of the fundamental by a factor of thirty and returns a value beyond the rigid-joint limit of the panel.
Condensing that residual flexibility onto the retained amplitudes rather than enlarging the basis to span it reproduces the full-order peak shift to $0.01$ percentage points for a hardening law and to $0.13$ for a fully slipped saturating one. How accurately it must be condensed is set by one spectral radius, since any error in the joint receptance is amplified at the solution by $(1-\rho)^{-1}$, which on this panel reaches twenty once the joints slip together.
All results are produced by an openly released implementation.
\end{abstract}


\section{Introduction}
\label{sec:intro}

Built-up aerospace structures are assemblies, and much of what makes them difficult to model lives in the joints. A fastened interface transmits shear by friction, and once the interface force exceeds what friction can carry, the two parts slip against each other. The resulting amplitude dependence is weak enough to be invisible in a low-level modal test and strong enough to matter at operating levels, which makes it a natural target for simulation and an awkward one for reduced-order modelling, because a localised nonlinearity interacts with a global basis in a way that a linear convergence study does not reveal.

The case considered here is a stiffened aluminium panel designed at the University of Sheffield and used in \citep{papatheou2008isma,papatheou2010pseudo} to study pseudo-fault novelty detection. The same physical panel was reused in \citep{battu2025transfer} to support a transfer-learning framework bridging finite-element predictions and experimental transmissibility features, and in \citep{hollins2026projection} as the first case study of a reduced-order-basis adaptation scheme. Across those studies the geometry, the instrumentation and the acquisition chain are identical, which makes the panel a useful common reference and its finite-element description worth writing down carefully.

Two bodies of work meet in this problem. Joint modelling in built-up structures has long recognised that a fastened interface is neither rigid nor linear, and that its microslip governs both the damping and the amplitude dependence of an assembly, which is why interface elements ranging from distributed springs to Iwan and Jenkins formulations are used in place of a bonded interface. Model updating, in its sensitivity form, provides the machinery for fixing the linear parameters of such a model against measured modal data, but it is a linear procedure and says nothing about which nonlinear description the joints should carry. Nonlinear model order reduction, in turn, is usually developed and tested on structures whose nonlinearity is distributed, geometric and smooth, such as the large-deflection behaviour of slender aerospace structures, where a modal basis that resolves the linear response also resolves the nonlinear one. A jointed panel is the awkward case for that assumption, because the nonlinearity is concentrated at a few hundred points that the global modes describe poorly, and the present study is an attempt to say precisely what goes wrong there and what fixes it.

This paper does three things with it. \Cref{sec:model,sec:updating} build an open shell finite-element model whose joints sit at the physical fastener positions rather than being smeared over a contact footprint, and calibrate it against the measured modes by a sensitivity method. \Cref{sec:nlmodel,sec:nlresults} make those fasteners nonlinear and bound the effect on the first three resonances, solving the resulting model twice over by routes that share nothing but the model matrices and the constitutive law, so that the effect is a property of the model and not of a solver. \Cref{sec:nrom} then applies a projection-based nonlinear model order reduction and identifies the quantity that actually sizes the basis of such a reduction, which turns out not to be the quantity a linear study would suggest, together with a reformulation that satisfies it without enlarging the basis.

A longer version of this study, with the complete verification suites, the mesh-convergence and fastener-stiffness sweeps and the full parameter tables, accompanies the released implementation.

\section{The panel, the measurements and the shell model}
\label{sec:model}

\subsection{Structure and data}

The structure is an aluminium panel of in-plane dimensions $750\times500\times3~\mathrm{mm}$, riveted to two C-channel ribs along the short edges (a $2''\times1''$ channel of depth $50~\mathrm{mm}$ with $25~\mathrm{mm}$ flanges) and bolted to two equal-leg angle-section stringers of length $700~\mathrm{mm}$ running parallel to the long edge (a $1''\times1''$ angle with $25~\mathrm{mm}$ web and leg). The drawing specifies the fastening, the ribs being secured with rivets at $30~\mathrm{mm}$ centres and the stringers with bolts at the same spacing, which gives one fastener row along the mid-line of each rib top flange, $16$ rivets each, and of each stringer leg, $23$ bolts each, so $78$ fasteners in total.

For the modal test the panel is suspended in a free--free configuration by extension springs and nylon lines, excited by a shaker driving a force transducer, and measured with fourteen unidirectional accelerometers bonded to the top sheet. A modal identification of the undamaged panel returns eighteen modes within the measurement band. The five retained here as the updating target are those tracked in the previous studies of the panel, at $24.70$, $69.99$, $104.31$, $117.56$ and $124.00~\mathrm{Hz}$. The fundamental is the first torsion mode, antisymmetric along the length with opposite sign on the two long edges, and the first bending mode about the short axis is the one at $104.31~\mathrm{Hz}$.

\subsection{Weak form and kinematics}

The continuum problem is the free vibration of a Reissner--Mindlin plate assembly comprising eleven shell substructures, one top plate, two stringer webs, two stringer legs, two rib webs and four rib flanges. Each substructure carries its own thickness $t$, density $\rho$ and effective Young's modulus $E_s = f_s E_0$, where $f_s$ is a substructure multiplier that the updating of \Cref{sec:updating} adjusts, with nominal constants $E_0 = 70~\mathrm{GPa}$, $\nu = 0.3$ and $\rho = 2700~\mathrm{kg/m^3}$. The weak form sought is, for all admissible test functions $(\vv,\bm{\psi})$,
\begin{equation}
  \int_{\Omega}\!\vsig(\vu,\vtheta_b):\veps(\vv,\bm{\psi})\,\mathrm{d}V
  \;=\;
  \omega^{2}\!\int_{\Omega}\!\rho\,\vu\!\cdot\!\vv\,\mathrm{d}V,
  \label{eq:weakform}
\end{equation}
where $\vu = (u,v,w)$ are the translations and $\vtheta_b = (\theta_x,\theta_y)$ the bending rotations.

The rotations $(\theta_x,\theta_y,\theta_z)$ are the components of the physical right-hand rotation vector of the shell normal, so that the in-plane displacement at height $z$ above the midsurface is $(u + z\theta_y,\, v - z\theta_x)$. With this convention the kinematic decomposition is
\begin{equation}
  \veps_m = \mathrm{sym}\,\nabla u_{xy}, \qquad
  \bm{\kappa} = (\partial_x\theta_y,\; -\partial_y\theta_x,\; \partial_y\theta_y - \partial_x\theta_x), \qquad
  \bm{\gamma} = (\partial_x w + \theta_y,\; \partial_y w - \theta_x),
  \label{eq:kinematics}
\end{equation}
for the membrane strain, the curvature and the transverse shear. The choice matters for a folded-shell assembly, because the nodal rotation triplet is rotated between the local frame of each substructure and the global frame by the same $3\times3$ matrix as the translations, which is only legitimate if the triplet is a genuine vector. A triplet such as $(-\theta_x,-\theta_y,+\theta_z)$ does not transform as one, and an element built on it passes every flat-plate test while failing the folded ones, for the reason given in \Cref{sec:verification}. The constitutive matrices for an isotropic plate are
\begin{equation}
  \mD_m = \frac{Et}{1-\nu^{2}}\begin{bmatrix} 1 & \nu & 0 \\ \nu & 1 & 0 \\ 0 & 0 & (1-\nu)/2\end{bmatrix},
  \quad
  \mD_b = \frac{Et^{3}}{12(1-\nu^{2})}\begin{bmatrix} 1 & \nu & 0 \\ \nu & 1 & 0 \\ 0 & 0 & (1-\nu)/2\end{bmatrix},
  \quad
  \mD_s = k_s G t\,\mI_2,
  \label{eq:constitutive}
\end{equation}
with $G = E/(2(1+\nu))$ and the standard shear-correction factor $k_s = 5/6$.

\subsection{The mixed-interpolation shell element}
\label{sec:mitc4}

Each substructure is meshed with the four-node quadrilateral flat-shell element of \citep{bathe1985mitc4,dvorkin1984continuum}. The element combines, in a single $24$-degree-of-freedom local stiffness, a plane-stress bilinear membrane with full $2\times2$ Gaussian integration, a Reissner--Mindlin bending part with the same integration, and a Hughes--Brezzi drilling penalty \citep{hughes1989plate} on $\theta_z$ that removes the singularity which would otherwise be present where several coplanar elements meet.

Naive integration of the shear energy on bilinear elements at thin-plate aspect ratios leads to severe shear locking, with a spurious stiffness that raises every flexural mode as the slenderness grows. The mixed-interpolation idea avoids this by computing the two covariant transverse shear strains not at the Gauss points but at the four mid-edge tying points $A = (0,-1)$, $B = (1,0)$, $C = (0,1)$ and $D = (-1,0)$, and reconstructing the shear field inside the element by
\begin{equation}
  \gamma_{rz}(r,s) = \tfrac{1}{2}(1-s)\,\gamma_{rz}^{A} + \tfrac{1}{2}(1+s)\,\gamma_{rz}^{C},
  \qquad
  \gamma_{sz}(r,s) = \tfrac{1}{2}(1-r)\,\gamma_{sz}^{D} + \tfrac{1}{2}(1+r)\,\gamma_{sz}^{B},
  \label{eq:mitc}
\end{equation}
followed by the covariant-to-Cartesian map through the inverse Jacobian at each Gauss point. \Cref{eq:mitc} is equivalent to enforcing the Kirchhoff thin-plate constraint at the four tying points rather than pointwise at every Gauss station, which eliminates locking while preserving the full Reissner--Mindlin response for thick plates.

Each element is built in a local frame and transformed to the global frame by the block-diagonal map built from the rotation $\mRot\in\R^{3\times3}$ whose rows are the two in-plane axes and the normal of the substructure plane,
\begin{equation}
  \mT = \mathrm{blkdiag}(\underbrace{\mRot,\mRot}_{\text{node }1},\,\underbrace{\mRot,\mRot}_{\text{node }2},\,\underbrace{\mRot,\mRot}_{\text{node }3},\,\underbrace{\mRot,\mRot}_{\text{node }4}),
  \qquad
  \mK^{(\mathrm{glob})} = \mT^{\mathsf{T}}\mK^{(\mathrm{loc})}\mT .
  \label{eq:transform}
\end{equation}
The drilling stabilisation is used in its consistent form,
\begin{equation}
  \Pi_{\mathrm{dril}} = \tfrac{1}{2}\,\alpha\,G\,t \int_{A}\Bigl(\theta_z - \tfrac{1}{2}\bigl(\partial_x v - \partial_y u\bigr)\Bigr)^{2}\mathrm{d}A ,
  \qquad \alpha = 0.03,
  \label{eq:drilling}
\end{equation}
integrated with the same rule. The penalty couples $\theta_z$ to the in-plane vorticity so that only genuine rigid rotations are energy-free. A penalty on $\theta_z$ alone, omitting the vorticity term, would leave a constant-drilling kernel on every coplanar substructure and render the shifted pencil exactly singular. The consistent mass carries $\rho t$ on the three translations and $\rho t^{3}/12$ rotary inertia on the two bending rotations.

\subsection{Fastener elements and assembly}
\label{sec:joints}

The plate midsurface is placed at $z = 0$, the horizontal stiffener pieces at $z = -t_{\mathrm{off}}$ with $t_{\mathrm{off}} = (t_{\mathrm{plate}} + t_{\mathrm{stiff}})/2$, so that the two midsurfaces are in contact through their half-thicknesses, and the vertical pieces hang below the horizontal ones. This reproduces the physical layering and avoids the spurious doubling of material that occurs when flanges and plate are meshed coplanar. Each stiffener is a single folded shell whose web and leg share the nodes along their common fold line.

Each fastener connects a plate node $a$ to the stiffener node $b$ directly below it. The two components touch at a contact point $c$ whose displacement follows from either side through the rigid offset arms $\vr_a = (0,0,-t_{\mathrm{plate}}/2)$ and $\vr_b = (0,0,+t_{\mathrm{stiff}}/2)$,
\begin{equation}
  \vu_c^{(a)} = \vu_a + \vtheta_a\times\vr_a, \qquad \vu_c^{(b)} = \vu_b + \vtheta_b\times\vr_b .
  \label{eq:contact}
\end{equation}
In the rigid fastener model, used for the reference results, the stiffener node is slaved to the plate node,
\begin{equation}
  \vtheta_b = \vtheta_a, \qquad \vu_b = \vu_a + \vtheta_a\times(0,0,-t_{\mathrm{off}}),
  \label{eq:rigidlink}
\end{equation}
imposed exactly by a multi-point-constraint transformation. In the compliant model a translational spring $k_t$ acts on the relative displacement of the two contact-point images and a rotational spring $k_r$ on the relative rotation,
\begin{equation}
  \Pi_{\mathrm{fast}} = \tfrac{1}{2}\,k_t\,\bigl\|\vu_c^{(a)}-\vu_c^{(b)}\bigr\|^{2} + \tfrac{1}{2}\,k_r\,\bigl\|\vtheta_a-\vtheta_b\bigr\|^{2},
  \label{eq:fastener}
\end{equation}
which adds a $12\times12$ block per fastener, the rigid model being recovered as $k_t,k_r\to\infty$. Unless stated otherwise the rotational stiffness is tied to the translational one through a washer radius $r_w = 5~\mathrm{mm}$, $k_r = k_t r_w^{2}/2$.

\begin{figure}[htbp]
  \centering
  \includegraphics[width=\linewidth]{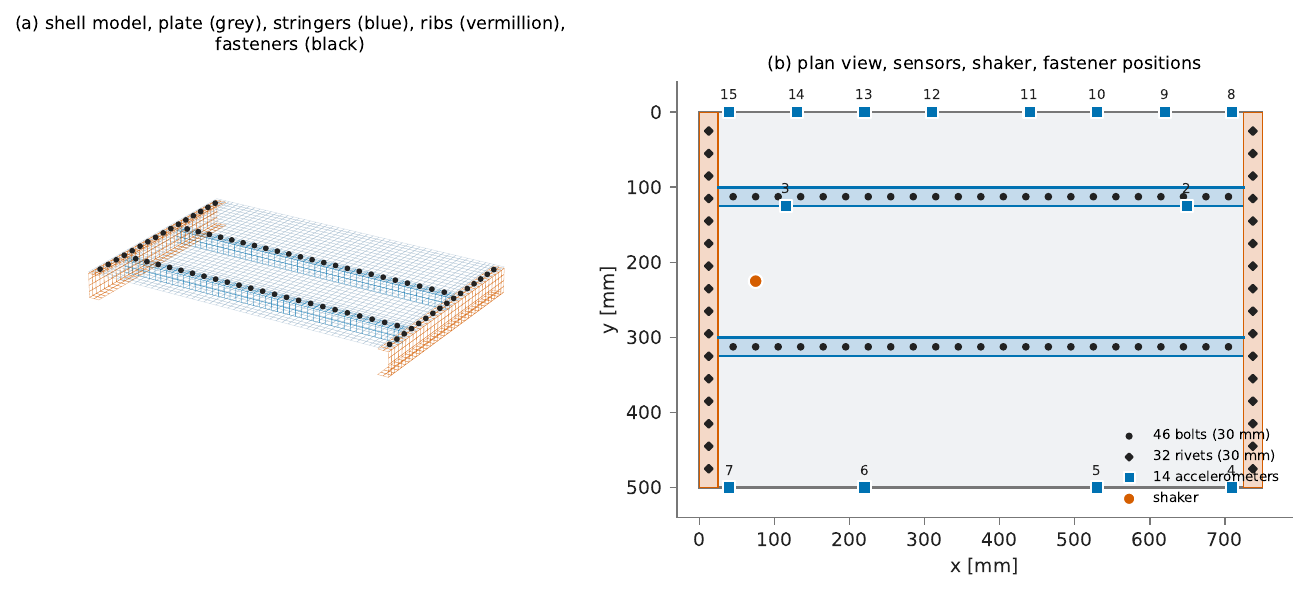}
  \caption{(a) Shell model at $h = 12.5~\mathrm{mm}$, plate (grey), stringers (blue), ribs (vermillion), and the $78$ fastener positions (black). (b) Plan view with the sensor numbering, the shaker and the fastener rows, $23$ bolts on the mid-line of each stringer leg and $16$ rivets on the mid-line of each rib top flange, all at $30~\mathrm{mm}$ centres.}
  \label{fig:geometry}
\end{figure}

Because the fastener count is a property of the structure and not of the mesh, both models are mesh-objective by construction, in contrast to a spring layer spread over every node inside a stiffener footprint, whose aggregate stiffness grows with the mesh. The mesh is forced to pass through every fastener position, every stiffener edge and every fold line. At the reference size $h = 12.5~\mathrm{mm}$ the model has $3{,}134$ nodes, $2{,}852$ elements and $18{,}804$ degrees of freedom, and the free-vibration problem
\begin{equation}
  \mK\,\vphi_k = \omega_k^{2}\,\mM\,\vphi_k
  \label{eq:eig}
\end{equation}
is solved by a shift-invert Arnoldi method \citep{lehoucq1998arpack} with the shift placed so that the six rigid-body modes and the lowest elastic modes are returned in a single solve. \Cref{fig:geometry} shows the model and the fastener rows.

\subsection{Verification}
\label{sec:verification}

The implementation is verified at the element, transformation, folded-assembly and system levels, \Cref{tab:verify}. The two folded benchmarks are the decisive ones for a stiffened panel, and the reason is worth stating. A single flat substructure is a similarity transform of itself, so a flat-plate benchmark cannot test \Cref{eq:transform} at all. An error in the rotation transformation is therefore invisible to every plate-only check while corrupting each fold line of a stiffened assembly. A free--free box beam gives its first bending frequency $2.2\%$ below the Euler--Bernoulli value, the residual being the shear deformation the beam formula ignores at a slenderness of $20$, and a free--free equal-leg angle gives its bending mode about the axis of symmetry at $184.7~\mathrm{Hz}$ against $185.2~\mathrm{Hz}$ from beam theory on the principal second moment of area. That mode bends both legs at once and can only be reproduced if the bending moment is transmitted correctly around the fold.

\begin{table}[htbp]
  \centering
  \caption{Verification of the shell finite-element pipeline.}
  \label{tab:verify}
  \small
  \begin{tabular}{l p{0.46\linewidth} r}
    \toprule
    level & test & result \\
    \midrule
    \multirow{3}{*}{element}
      & zero-energy modes of a single element & exactly $6$ \\
      & rigid rotation about $x$, $y$, $z$ & $<10^{-17}$~J on all three axes \\
      & simply supported square plate vs Navier & $+0.4$, $+1.5$, $+1.6\%$ (modes 1--3) \\
    \midrule
    \multirow{2}{*}{transformation}
      & free strip, in-plane mesh vs vertical mesh & $2.5\times10^{-10}$ relative \\
      & the same strip vs analytical thin-strip fundamental & $34.92$ vs $34.89~\mathrm{Hz}$ \\
    \midrule
    \multirow{2}{*}{folded assembly}
      & box beam, first bending vs Euler--Bernoulli $370.1~\mathrm{Hz}$ & $362.1~\mathrm{Hz}$, $-2.2\%$ \\
      & angle section, bending about the symmetry axis vs $185.2~\mathrm{Hz}$ & $184.7~\mathrm{Hz}$, $-0.3\%$ \\
    \midrule
    system
      & bare plate, first elastic (torsion) mode & $26.27~\mathrm{Hz}$ \\
    \bottomrule
  \end{tabular}
\end{table}

\section{Model updating}
\label{sec:updating}

\subsection{Formulation}

The updating follows the classical sensitivity-based formulation \citep{mottershead1993modelupdating,friswell1995model,mottershead2011sensitivity}. With rigid fasteners the parameter vector holds the six Young's-modulus multipliers of the substructure groups,
\begin{equation}
  \vtheta = \bigl(f_{\mathrm{plate}},\; f_{\mathrm{str,web}},\; f_{\mathrm{str,leg}},\; f_{\mathrm{rib1,web}},\; f_{\mathrm{rib2,web}},\; f_{\mathrm{rib,fl}}\bigr)\in\R^{6},
  \label{eq:theta}
\end{equation}
and with compliant fasteners the two logarithmic stiffnesses $\log_{10}k_b$ and $\log_{10}k_v$ are prepended. The residual carries both frequency and mode-shape information,
\begin{equation}
  \vr(\vtheta)
  = \Bigl[\,\underbrace{(f_k^{\,\mathrm{FE}}(\vtheta)-f_k^{\,\mathrm{m}})/f_k^{\,\mathrm{m}}}_{k=1,\dots,5}\;\;\Big|\;\;
  \underbrace{w_{\mathrm{MAC}}\bigl(1-\mathrm{MAC}(\vphi_k^{\,\mathrm{FE}},\vphi_k^{\,\mathrm{m}})\bigr)}_{k=1,\dots,5}\,\Bigr]^{\mathsf{T}},
  \label{eq:residual}
\end{equation}
where $\vphi_k^{\,\mathrm{FE}}$ is the finite-element mode evaluated at the fourteen accelerometer positions through a sensor-extraction matrix, and the modal assurance criterion \citep{allemang1982mac} is
\begin{equation}
  \mathrm{MAC}(\vphi^{\,\mathrm{FE}},\vphi^{\,\mathrm{m}})
  = \frac{\bigl|\vphi^{\,\mathrm{FE}\,*}\vphi^{\,\mathrm{m}}\bigr|^{2}}
         {(\vphi^{\,\mathrm{FE}\,*}\vphi^{\,\mathrm{FE}})(\vphi^{\,\mathrm{m}\,*}\vphi^{\,\mathrm{m}})} .
  \label{eq:mac}
\end{equation}
The weight $w_{\mathrm{MAC}} = 0.5$ balances the two residual blocks, giving ten residuals against six free parameters. With the sensitivity matrix $\mS_{ij} = \partial r_i/\partial\theta_j$ approximated by forward differences, each Levenberg--Marquardt iteration \citep{levenberg1944method,marquardt1963algorithm} solves
\begin{equation}
  (\mS^{\mathsf{T}}\mS + \lambda\mI)\,\Delta\vtheta = -\,\mS^{\mathsf{T}}\vr(\vtheta),
  \qquad
  \vtheta\,\leftarrow\,\Pi_{[\vtheta_{\min},\vtheta_{\max}]}\bigl(\vtheta + \Delta\vtheta\bigr),
  \label{eq:lm}
\end{equation}
with $\Pi$ the projection onto the box constraints and $\lambda$ decreased when a step is accepted and increased when it is rejected. Two conventions are admissible for pairing finite-element and measured modes, the order convention and the best-MAC convention. They differ only when the finite-element order does not match the measured one, which is the case here for measured modes 4 and 5, so the best-MAC convention is used and the order convention is run as a check.

\subsection{The calibrated model}

\Cref{tab:calib} gives the result. The uncalibrated model already finds all five shapes with modal-assurance values of $0.85$ or better, at frequency errors of $0.7$ to $10.1\%$, which is the same quality as the initial model of \citep{hollins2026projection}. The iteration converges in five accepted steps from the nominal model to an interior optimum,
\begin{equation*}
  \vtheta^{*} = (1.164,\;0.932,\;1.018,\;1.109,\;0.898,\;1.014),
\end{equation*}
an effective plate modulus of $81.5~\mathrm{GPa}$ with stiffener moduli within about $10\%$ of nominal, and brings the errors to $1.1$ to $4.3\%$. All six parameters are interior to the $\pm20\%$ box, the same vector is reached from a much wider box and with compliant fasteners, and an eight-mesh sweep to $300{,}612$ degrees of freedom leaves a mesh bias below $3\%$.

Two features of the residual are worth recording. It is systematic rather than random, the fundamental and measured mode 5 remaining low while modes 2 to 4 are slightly high, and no combination of the six moduli within physical bounds removes it. And the calibration brings finite-element modes 4 and 5 to within $1~\mathrm{Hz}$ of each other where the measured pair is $6.4~\mathrm{Hz}$ apart, which is why those two carry the lowest modal-assurance values of the set.

A sweep of the fastener stiffness over six orders of magnitude shows that the torsion fundamental is present at every stiffness and that its position is governed by the rotational coupling of the fastener line, which lifts it by $15\%$ relative to translation-only fasteners. Fastener stiffnesses in the physically plausible range are already in the saturated regime, which is why the rigid-link model is an adequate reference and why the calibration with compliant fasteners leaves the stiffnesses unchanged.

\begin{figure}[htbp]
  \centering
  \includegraphics[width=0.72\linewidth]{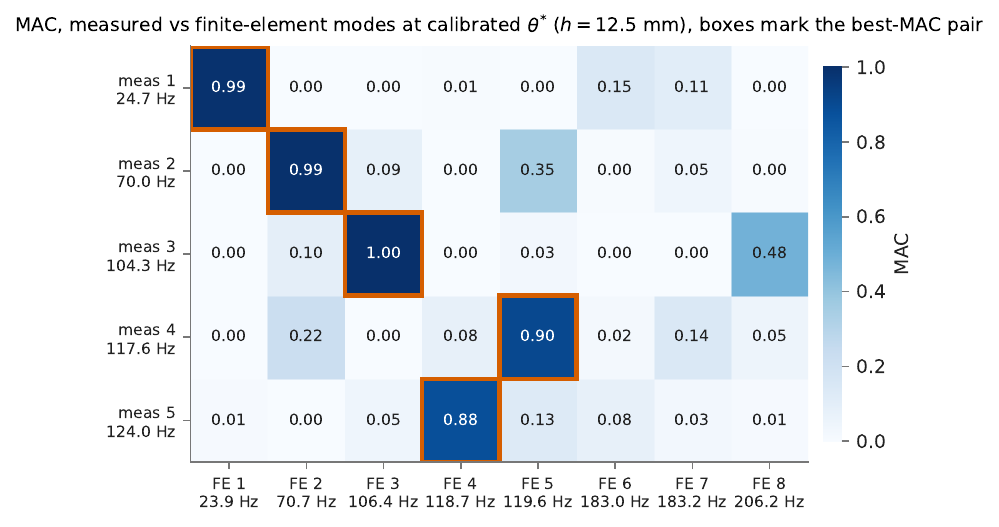}
  \caption{Modal assurance matrix between the five measured modes and the first eight finite-element elastic modes at the calibrated parameters, boxes marking the best pair of each measured mode. The matrix is diagonal up to the interchange of modes 4 and 5.}
  \label{fig:mac}
\end{figure}

\begin{table}[htbp]
  \centering
  \caption{Uncalibrated and calibrated models at $h = 12.5~\mathrm{mm}$ against the five tracked measured modes, under best-MAC pairing. Rigid fasteners throughout.}
  \label{tab:calib}
  \small
  \begin{tabular}{c r r r r r r}
    \toprule
    & & \multicolumn{2}{c}{uncalibrated} & \multicolumn{2}{c}{calibrated} & \\
    \cmidrule(lr){3-4}\cmidrule(lr){5-6}
    mode & measured [Hz] & FE [Hz] & err [\%] & FE [Hz] & err [\%] & MAC \\
    \midrule
    1 & 24.70  & 22.51  & $-8.9$  & 23.86  & $-3.4$ & 0.99 \\
    2 & 69.99  & 67.38  & $-3.7$  & 70.73  & $+1.1$ & 0.99 \\
    3 & 104.31 & 103.24 & $-1.0$  & 106.36 & $+2.0$ & 1.00 \\
    4 & 117.56 & 118.43 & $+0.7$  & 119.59 & $+1.7$ & 0.90 \\
    5 & 124.00 & 111.45 & $-10.1$ & 118.70 & $-4.3$ & 0.88 \\
    \bottomrule
  \end{tabular}
\end{table}

\subsection{Mesh refinement and the fastener stiffness}
\label{sec:sweeps}

Two sweeps bound the trust that can be placed in the calibrated parameters, and both are run at the calibrated point rather than at the nominal one.

\begin{figure}[htbp]
  \centering
  \includegraphics[width=\linewidth]{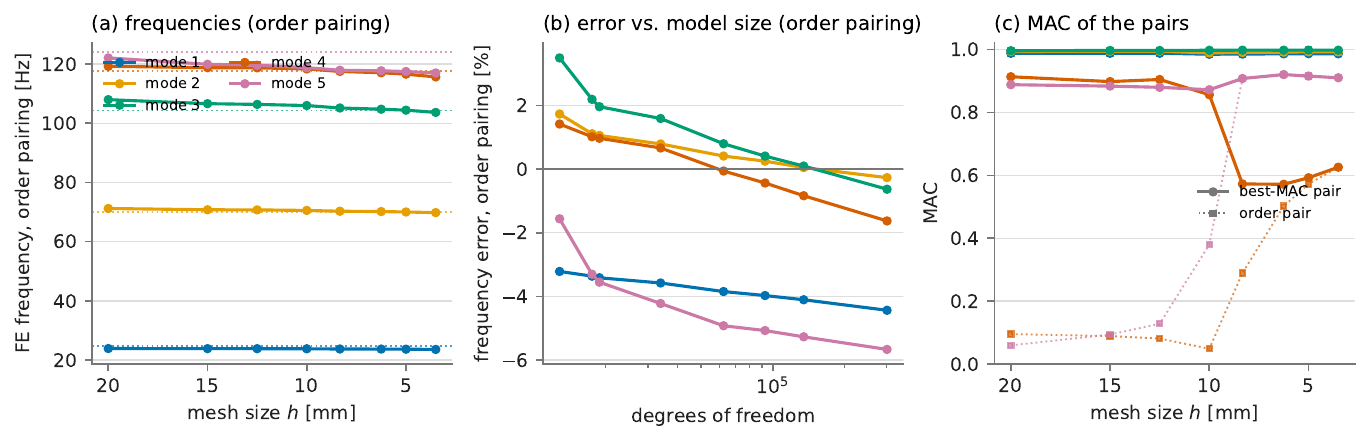}
  \caption{Mesh convergence of the calibrated model across eight mesh sizes, (a) finite-element frequencies with the measured values dotted, (b) frequency error against model size, (c) modal-assurance value of each pair. The fastener count is fixed at $78$ for every mesh. The dip of the mode-4 value at the finer meshes is the hybridisation of the nearly coincident finite-element modes 4 and 5.}
  \label{fig:convergence}
\end{figure}

The first refines the mesh across eight sizes from $h = 20$ to $3.5~\mathrm{mm}$, that is from $12{,}804$ to $300{,}612$ degrees of freedom, with the same $78$ fasteners at every mesh, \Cref{fig:convergence}. The five frequencies converge monotonically from above, as expected of a displacement-based element, and from $h = 12.5$ to $3.5~\mathrm{mm}$ the errors move from $(-3.4, +1.1, +2.0, +1.0, -3.6)\%$ to $(-4.4, -0.3, -0.6, -1.6, -5.7)\%$, a mesh bias of $1$ to $2.6$ percentage points at the calibration mesh, with the two finest meshes differing by less than $1\%$. The shapes of measured modes 1, 2, 3 and 5 keep a modal-assurance value of $0.91$ or better throughout. The exception is the pair the calibration brought within $1~\mathrm{Hz}$ of each other, which hybridises and exchanges order between $h = 10$ and $8.33~\mathrm{mm}$, so that the modal-assurance value of measured mode 4 dips while the frequencies of the pair stay within $2\%$. Repeating the whole calibration from the nominal model at $h = 6.25~\mathrm{mm}$ returns a parameter vector within $7\%$ of the reference one in every modulus, so the calibrated moduli are not mesh artefacts.

\begin{figure}[htbp]
  \centering
  \includegraphics[width=\linewidth]{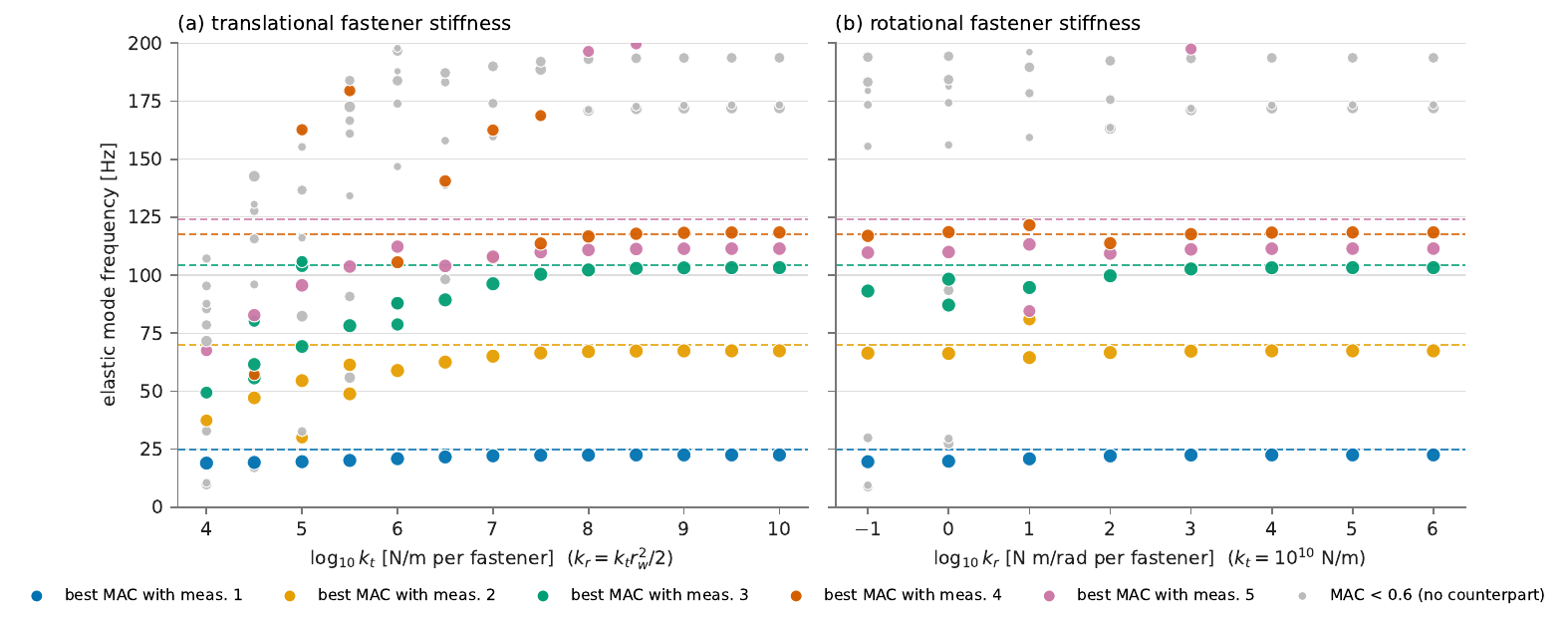}
  \caption{Elastic spectrum of the model against (a) the translational fastener stiffness and (b) the rotational fastener stiffness. Each mode is coloured by the measured mode with which it correlates best, grey markers having no counterpart. Dashed lines mark the five measured frequencies. The torsion fundamental is present at every stiffness.}
  \label{fig:fastener}
\end{figure}

The second sweeps the compliant-fastener stiffness over six orders of magnitude, \Cref{fig:fastener}. The torsion fundamental is present at every stiffness, with a modal-assurance value of $0.99$ or better, moving only between $19.0~\mathrm{Hz}$ with the softest fasteners and $22.5~\mathrm{Hz}$ in the rigid limit. Its position is governed by the rotational coupling rather than the translational one, since with translation-only fasteners it sits at $19.6~\mathrm{Hz}$, $21\%$ below the measurement, and it reaches its rigid-fastener value once the rotational stiffness exceeds about $10^{3}~\mathrm{N\,m/rad}$. Modes 2 to 5 are within $1.5\%$ of their rigid-fastener values for translational stiffnesses above $10^{8}~\mathrm{N/m}$ and still $3$ to $9\%$ below them at $10^{7}~\mathrm{N/m}$, and below $10^{5}~\mathrm{N/m}$ the band under $120~\mathrm{Hz}$ fills with stiffener-dominated modes that have no measured counterpart. Semi-empirical estimates place a single fastener of this size in $3~\mathrm{mm}$ aluminium at $10^{7}$ to $10^{8}~\mathrm{N/m}$, at the upper end of the transition, which is why the calibration with compliant fasteners finds no sensitivity there and why the rigid-link model is used as the reference.

\section{The nonlinear joint model and how it is solved}
\label{sec:nlmodel}

\subsection{Equation of motion and joint laws}

The calibrated model is kept exactly as it is and the fastener elements are made nonlinear, so that the linear model can be compared, at equal mesh, mass and damping, with nonlinear models that differ from it only in the constitutive law of the joints. The equation of motion is
\begin{equation}
  \mM \ddot{\vu} + \mC\dot{\vu} + \mK \vu + \mB^{\mathsf{T}}\!\left[\bm{g}(\mB\vu, \bm{p}) - \bm{K}_{\!j}\,\mB\vu\right] = \vf(t),
  \label{eq:nleom}
\end{equation}
where $\mK$ and $\mM$ are the calibrated matrices with compliant fasteners, $\mC = \alpha\mM + \beta\mK$ is Rayleigh damping fixed by the measured modal damping, $\mB$ extracts the joint deformations, $\bm{p}$ collects the internal states of the hysteretic elements, and $\bm{K}_{\!j}$ is the diagonal matrix of linear joint stiffnesses. The bracket is the correction of the fastener force with respect to the linear fastener already contained in $\mK$, so a linear law reproduces the linear model identically, which is used below as a verification test.

Each fastener carries two isotropic two-component elements, the relative in-plane translation of the two contact points, the slip, with stiffness $k_t$, and the relative bending rotation of the two nodes with stiffness $k_r$. The relative normal translation and the drilling rotation are left linear. This gives $156$ nonlinear elements and $\mB \in \R^{312\times 18{,}804}$. The split matters, because the covered joint deformation carries $0.41$, $0.96$ and $1.82\%$ of the strain energy of the first three modes, and for the torsion fundamental $88\%$ of that is rotational rather than translational, so a model that made only the slip nonlinear would barely register on mode 1.

\begin{figure}[htbp]
  \centering
  \includegraphics[width=\linewidth]{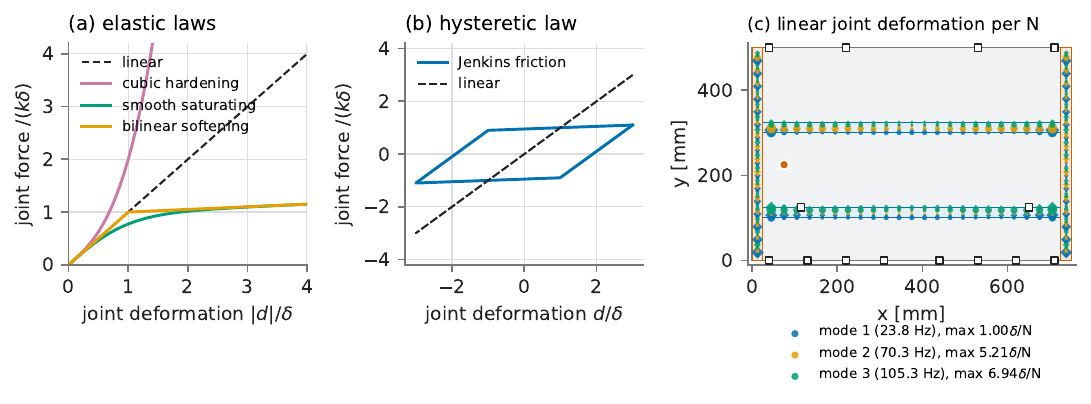}
  \caption{(a) The three elastic joint laws and (b) the hysteretic one, all with the same initial stiffness $k$ and knee deformation $\delta$, and the same retained fraction $\varrho = 0.05$ beyond saturation. (c) Joint deformation of the linear model per newton of shaker force at each of the first three resonances, in units of the knee. The knee is defined so that the most loaded element reaches it at $1~\mathrm{N}$ at the fundamental.}
  \label{fig:nllaws}
\end{figure}

Four laws are compared, all sharing the same initial stiffness $k$ and knee deformation $\delta$, \Cref{fig:nllaws},
\begin{align}
  \text{cubic hardening} \quad & s(r) = k r + (k/\delta^{2})\,r^{3}, \label{eq:lawcubic}\\
  \text{smooth saturating} \quad & s(r) = \varrho\,k r + (1-\varrho)\,k\delta\tanh(r/\delta), \label{eq:lawtanh}\\
  \text{bilinear softening} \quad & s'(r) = k \ \ (r\le\delta), \qquad s'(r) = \varrho k \ \ (r>\delta), \label{eq:lawbilin}\\
  \text{Jenkins friction} \quad & \bm{g} = \varrho k \vd + (1-\varrho)\,\bm{g}_{J}, \label{eq:lawjenkins}
\end{align}
with $\bm{g}_{J}$ an elastic-Coulomb element of stiffness $k$ and slip force $k\delta$, integrated by radial return with its consistent tangent. The retained fraction $\varrho = 0.05$ is what a slipped fastener still carries through its shank in bearing. It is not cosmetic, since with $\varrho = 0$ the saturated joint leaves the stiffeners free to slide on the plate, the model acquires twelve zero-energy mechanisms beyond the six rigid-body modes, and the computed response is dominated by that artefact.

The knee is set once from the linear model, $\delta_s$ and $\delta_r$ being the largest joint slip and joint rotation produced anywhere in the panel by $1~\mathrm{N}$ of shaker force at the fundamental resonance, which gives $\delta_s = 17.3~\mathrm{nm}$ and $\delta_r = 9.29~\mu\mathrm{rad}$, a slip force of $1.73~\mathrm{N}$ per fastener. Excitation levels of $1$, $3$, $10$ and $30~\mathrm{N}$ then span the range from an essentially linear response to one in which almost every joint element is driven well past its knee. Because the joint deformation per newton is $5.2$ and $6.9$ times larger at modes 2 and 3 than at mode 1, the same joint produces very different degrees of nonlinearity in the three modes.

\subsection{Two independent solution paths}
\label{sec:nlsolution}

The nonlinear response is computed twice, by solvers that have nothing in common but the model matrices, the constitutive law and the sparse-factorisation utility, one working in the time domain on every degree of freedom, the other in the frequency domain on the joints alone.

The first is direct time integration of \Cref{eq:nleom} on all $18{,}804$ degrees of freedom, Newmark with the average-acceleration parameters and a damped Newton iteration at every step. The nonlinear term is localised, so the tangent is the constant matrix $\mK + a_0\mM + a_1\mC$ plus a low-rank update, and one sparse factorisation serves the whole run through the Woodbury identity, which on this model replaces a $3.7~\mathrm{s}$ factorisation per iteration by a $0.18~\mathrm{s}$ solve. The scheme satisfies an exact discrete energy identity, monitored throughout.

The second is harmonic balance with exact dynamic condensation onto the joint degrees of freedom. Writing the response as $\vu(t) = \sum_h \Re\{\bm{U}_h e^{\mathrm{i}h\omega t}\}$ and eliminating $\bm{U}_h$ gives a system in the $312$ joint deformations alone,
\begin{equation}
  \vd_h = F_0\,\bm{e}_1\,\delta_{h1} - \bm{R}_h\,\bm{Q}_h(\vd), \qquad
  \bm{R}_h = \mB\,\bm{Z}(h\omega)^{-1}\mB^{\mathsf{T}},
  \label{eq:condensed}
\end{equation}
with $\bm{Z}(\omega) = \mK - \omega^{2}\mM + \mathrm{i}\omega\mC$, $\bm{e}_1$ the joint deformation produced by a unit shaker force, and $\bm{Q}_h$ the Fourier coefficients of the nonlinear correction force, evaluated by an alternating frequency-time scheme with its exact Jacobian. For the friction law the sensitivity of every sample is propagated through the return-mapping recursion. The condensation is exact, $\bm{R}_h$ being the receptance of the full model between the joint degrees of freedom rather than of a reduced one.

Evaluating $\bm{R}_h$ by direct complex solves would cost $313$ sparse solves per frequency and per harmonic. With Rayleigh damping this is unnecessary, because the dynamic stiffness factorises,
\begin{equation}
  \bm{Z}(\omega) = (1 + \mathrm{i}\omega\beta)\left(\mK - \mu\mM\right), \qquad
  \mu(\omega) = \frac{\omega^{2} - \mathrm{i}\omega\alpha}{1 + \mathrm{i}\omega\beta},
  \label{eq:rayleigh}
\end{equation}
so that in the real mass-orthonormal modes of the undamped model $\bm{Z}^{-1}$ is a sum over modes with a single scalar denominator each. Truncating that sum to $150$ modes and correcting the remainder by its value and slope at one real reference point, at the cost of a single real factorisation, reproduces the direct complex solves to $2.7\times10^{-9}$ at $23.8~\mathrm{Hz}$ and $6.6\times10^{-10}$ at $71.4~\mathrm{Hz}$, and makes any frequency essentially free to evaluate. The correction is not optional, the $150$ retained modes on their own being $94\%$ in error, because the joint receptance is dominated by the local static flexibility of the fastener region rather than by the resonances. This observation returns, in a sharper form, in \Cref{sec:nrombasis}.

A softening joint bends the resonance far enough that the response becomes multivalued and a stepped sweep fails at the turning point. The branches are therefore traced by pseudo-arclength continuation, with the excitation frequency as an unknown and the frequency derivative of \Cref{eq:condensed} assembled analytically, so that folds are passed and the unstable middle branch is obtained as well.

\subsection{Verification}

\Cref{tab:nlverify} lists the verification of this machinery. Every entry is a relative deviation from an independent reference, an analytical result where one exists, an independently coded implementation of the same quantity, or a property that holds exactly by construction. The two entries that matter most are the agreement of the two solution paths on the full model, $5.5\times10^{-4}$ in the fundamental amplitude and $7.0\times10^{-4}$ in the energy dissipated per cycle, since those two share no solver code, and the reproduction of the linear model by a linear law, which tests the whole nonlinear pipeline against the model of \Cref{sec:updating}.

\begin{figure}[htbp]
  \centering
  \includegraphics[width=\linewidth]{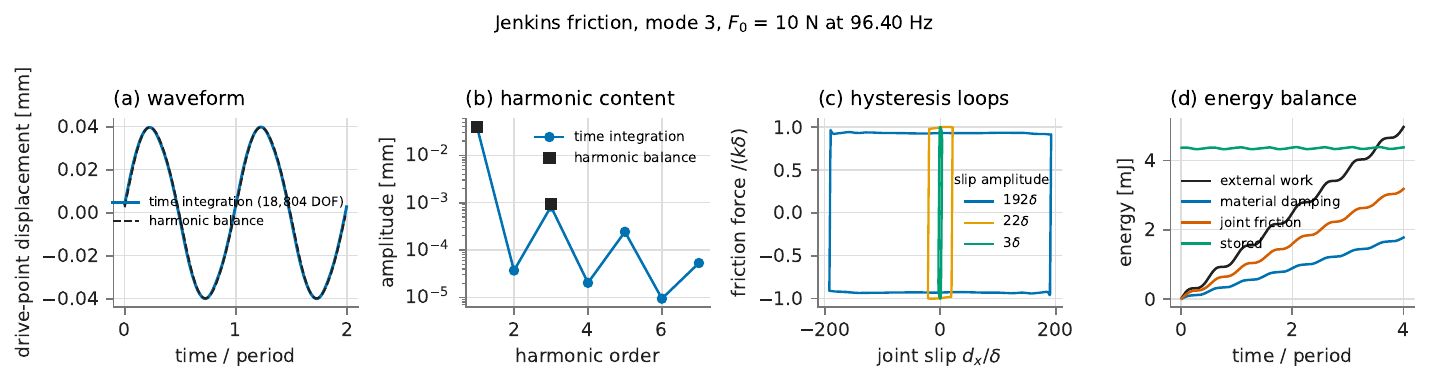}
  \caption{The two solution paths at one strongly nonlinear operating point, mode 3 under friction at $10~\mathrm{N}$, (a) steady-state waveform, (b) harmonic content, (c) joint hysteresis loops, (d) energy balance. The time integration runs on all $18{,}804$ degrees of freedom and the harmonic balance on the $312$ joint unknowns.}
  \label{fig:timedomain}
\end{figure}

\Cref{fig:timedomain} closes the loop on the two paths at one strongly nonlinear operating point. The time-integrated waveform of the full model and the harmonic-balance waveform lie on top of each other, the harmonic content agrees, the joint hysteresis loops are traced directly, and the energy accounting closes, with the external work absorbed by material damping and joint friction in the ratio both methods predict. The distortion is modest at these levels, at most $5.1\%$ of the fundamental at the resonance peaks, which is why a two-harmonic set suffices, and the nonlinearity shows itself in the shift and the loss of the peak rather than in a visibly distorted waveform.

One feature of the traced branches is worth recording because it is easy to mistake for a solver failure. Narrow spikes a few hertz wide appear at frequencies where three times the excitation frequency meets a mode of the panel, and these are $1{:}3$ superharmonic resonances, carrying the largest third-harmonic content of the whole study. They are the reason the harmonic set has to be checked rather than assumed, since the seventh harmonic changes the fundamental at those points by more than it does anywhere else.

\begin{table}[htbp]
  \centering
  \caption{Verification of the nonlinear full-order machinery. Every entry is a relative deviation from an independent reference.}
  \label{tab:nlverify}
  \small
  \begin{tabular}{l p{0.50\linewidth} r}
    \toprule
    level & check & deviation \\
    \midrule
    \multirow{2}{*}{element}
      & law tangents vs central finite differences, all four laws & $6.0\times10^{-11}$ \\
      & hysteretic Jacobian vs finite differences & $1.0\times10^{-9}$ \\
    \midrule
    \multirow{4}{*}{single degree of freedom}
      & Duffing, Newmark vs an adaptive reference integrator & $1.8\times10^{-4}$ \\
      & Newmark convergence order in $\Delta t$, deviation from 2 & $1.4\times10^{-4}$ \\
      & Jenkins hysteresis loop area vs $4F_s(A-F_s/k)$ & $2.2\times10^{-4}$ \\
      & Duffing, harmonic balance vs the analytical relation & $4.3\times10^{-11}$ \\
    \midrule
    \multirow{3}{*}{full-order model}
      & Woodbury tangent solve vs the assembled tangent & $1.2\times10^{-14}$ \\
      & zero nonlinearity reproduces the linear solution & $2.0\times10^{-11}$ \\
      & condensed receptance vs direct solves at $23.80$~Hz & $2.7\times10^{-9}$ \\
    \midrule
    \multirow{2}{*}{nonlinear response}
      & friction law, harmonic balance vs time integration & $5.5\times10^{-4}$ \\
      & the same, energy dissipated per cycle & $7.0\times10^{-4}$ \\
    \midrule
    continuation
      & traced branch through a fold vs the analytical relation & $2.5\times10^{-10}$ \\
    \bottomrule
  \end{tabular}
\end{table}

\section{What joint nonlinearity does to the panel}
\label{sec:nlresults}

\begin{figure}[htbp]
  \centering
  \includegraphics[width=\linewidth]{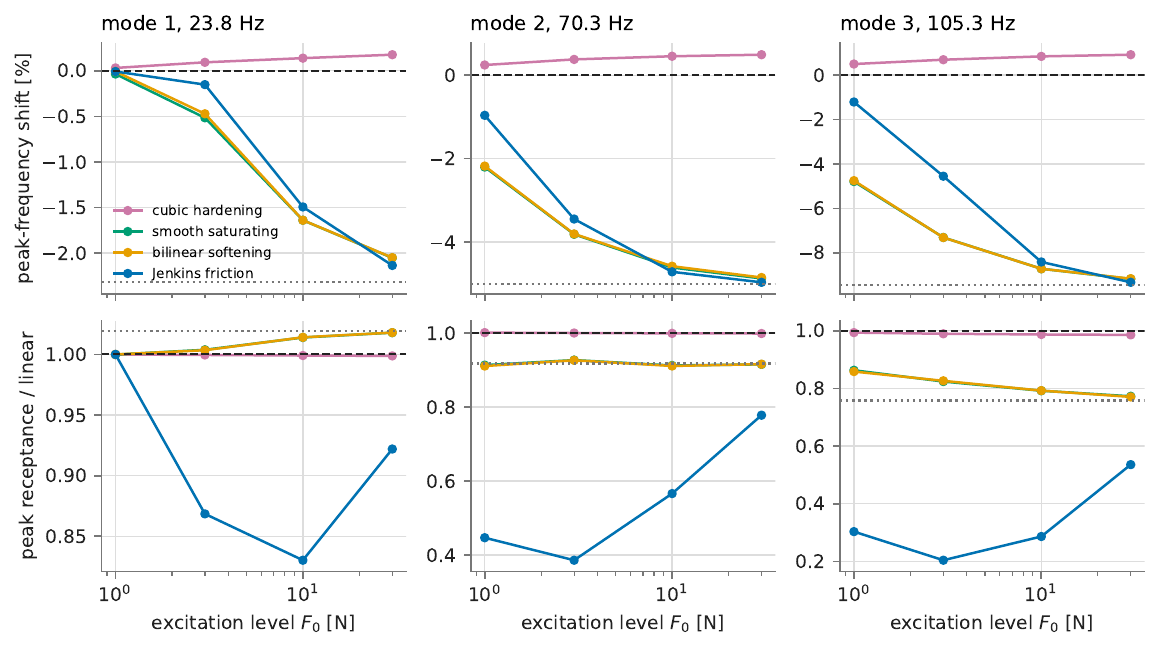}
  \caption{Level dependence of the resonance peak for the four joint laws and the first three modes, shift of the peak frequency from the linear model (top) and ratio of the peak receptance to the linear one (bottom). Dashed lines, the linear model. Dotted lines, the fully saturated bound.}
  \label{fig:backbone}
\end{figure}

The main result is an asymmetry, shown in \Cref{fig:backbone} and summarised in \Cref{tab:nlsummary}. Hardening the joints does almost nothing, the cubic law shifting the peak by less than $+1\%$ even when the most loaded element is driven to six times its knee, because a joint already nearly rigid compared with the sheet cannot stiffen the structure further and the strain energy simply stays in the plate. Softening or slipping them does a great deal, because removing the constraint changes the structure. The saturating and bilinear laws move the peak by $-2.1$, $-4.9$ and $-9.2\%$ on the first three modes, and the friction law removes up to $80\%$ of the resonant peak at intermediate levels before it recovers at high level, once the joints slip through most of the cycle.

\begin{figure}[htbp]
  \centering
  \includegraphics[width=\linewidth]{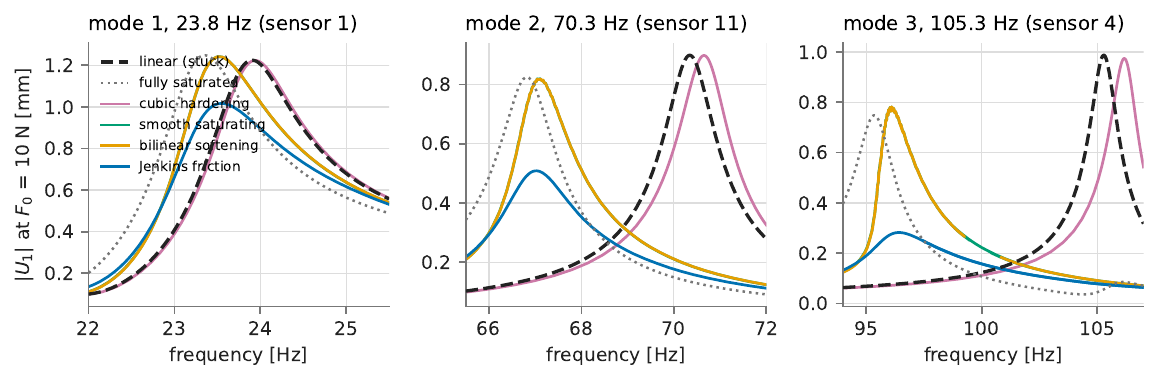}
  \caption{Linear against nonlinear full-order response at $10~\mathrm{N}$, at the sensor with the largest response in each band. Dashed, the linear model. Dotted, the fully saturated bound. Coloured, the four nonlinear models, which differ from the linear one only in the constitutive law of the $78$ fasteners.}
  \label{fig:frflaws}
\end{figure}

\Cref{fig:frflaws} shows the response curves themselves at an intermediate level, which makes the asymmetry visible before any peak is extracted. The cubic branch is indistinguishable from the linear one at the scale of the plot. The two softening branches have moved left and lost amplitude, and the friction branch has lost the most amplitude of all while moving least, because friction removes energy as well as stiffness. All of them lie between the linear model and the fully saturated bound.

Two bounds make the numbers interpretable. The shifts approach, to within $0.3\%$, a fully saturated limit computed from a separate eigenvalue problem in which every joint element retains only $\varrho k$ and which carries the same damping matrix as the nonlinear model. On the channel plotted here every nonlinear branch approaches that bound from above without crossing it. And the shifts remain proportional to the fraction of modal strain energy the joints carry, which never exceeds $2\%$ of the total, while being eleven times larger than a first-order perturbation of that energy would predict. The effect is therefore not a perturbation of the linear model, but it is bounded, and the level range in which it would first become visible is fixed.

These are numerical experiments on a calibrated model, not a measurement of the panel. The knee and the retained fraction are chosen rather than identified, since the available dataset is a single low-level modal identification which by construction contains no information about amplitude dependence. What the study fixes is the machinery, the bounds and the level range, and the obvious continuation is experimental, since a stepped-sine test of the same panel at several levels would decide directly whether its joints behave in this way.

\begin{figure}[htbp]
  \centering
  \includegraphics[width=\linewidth]{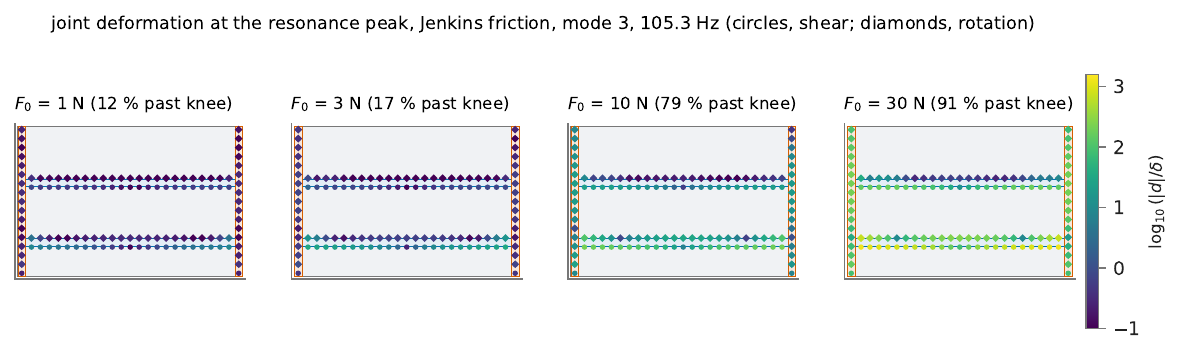}
  \caption{Where the panel is nonlinear, joint deformation at the resonance peak of mode 3 under Jenkins friction, in units of the knee, at the four excitation levels. Marker area is proportional to the deformation and the fraction of elements past the knee is quoted for each level.}
  \label{fig:slipmap}
\end{figure}

Where the panel is nonlinear is as informative as by how much, \Cref{fig:slipmap}. At $1~\mathrm{N}$ the fundamental is still entirely stuck and its response is indistinguishable from linear, while mode 3, whose joints see seven times more deformation per newton, has already lost two thirds of its peak. As the level rises the fraction of joint elements past their knee grows from a handful clustered at the most loaded fastener rows to almost the whole population, and the peak of each mode falls to a minimum and then recovers, since at $30~\mathrm{N}$ the joints slip through most of the cycle and the structure spends most of its period in the fully saturated configuration rather than passing through it.

\begin{table}[htbp]
  \centering
  \caption{Peak-frequency shift from the linear model at $30~\mathrm{N}$, for the first three modes, together with the fully saturated bound.}
  \label{tab:nlsummary}
  \small
  \begin{tabular}{l r r r}
    \toprule
    joint law & mode 1 [\%] & mode 2 [\%] & mode 3 [\%] \\
    \midrule
    cubic hardening      & $+0.17$ & $+0.49$ & $+0.92$ \\
    smooth saturating    & $-2.06$ & $-4.87$ & $-9.17$ \\
    bilinear softening   & $-2.05$ & $-4.85$ & $-9.15$ \\
    Jenkins friction     & $-2.14$ & $-4.97$ & $-9.32$ \\
    \midrule
    fully saturated bound & $-2.31$ & $-5.01$ & $-9.43$ \\
    \bottomrule
  \end{tabular}
\end{table}

\section{Nonlinear model order reduction}
\label{sec:nrom}

\subsection{Descriptor form, Taylor expansion and projection}
\label{sec:nromform}

The reduction applied here is the projection-based nonlinear model order reduction developed for very flexible aircraft, in which the response is projected onto a small basis and the nonlinear restoring force is carried by a Taylor expansion about the equilibrium. It was built for coupled aeroelastic and flight-dynamic systems, where it reduces a geometrically nonlinear beam together with its unsteady aerodynamics to a handful of states at speedups of two to three orders of magnitude \citep{tantaroudas2026nmor}, and it has since served as the plant for gust and control studies on the same class of vehicle, rapid worst-case gust identification \citep{tantaroudas2026gust}, robust gust load alleviation \citep{tantaroudas2026hinf}, and nonlinear flutter suppression carried from simulation through to wind tunnel testing \citep{daronch2014flutter}. Those applications share a feature that the present structure does not, since their nonlinearity is geometric, distributed along a slender member and smooth, so that a basis resolving the linear response resolves the nonlinear one as well. A jointed panel is the opposite case, and the rest of this section is an attempt to say precisely what that costs and what can be done about it.

The full-order model of \Cref{eq:nleom} is written in first-order form. Because the consistent mass matrix carries no inertia on the drilling degrees of freedom it is singular, so the first-order system is a descriptor system rather than an ordinary differential equation,
\begin{equation}
  \bm{E}\,\dot{\vw} \,=\, \bm{A}\,\vw \,+\, \bm{N}(\vw) \,+\, \vf(t), \qquad
  \vw = \begin{Bmatrix} \vu \\ \dot{\vu} \end{Bmatrix}, \quad
  \bm{E} = \begin{bmatrix} \mI & \bm{0} \\ \bm{0} & \mM \end{bmatrix}, \quad
  \bm{A} = \begin{bmatrix} \bm{0} & \mI \\ -\mK & -\mC \end{bmatrix},
  \label{eq:descriptor}
\end{equation}
with $\bm{N}(\vw) = \{\bm{0};\,-\mB^{\mathsf{T}}\vs(\mB\vu)\}$ and $\vs(\vd) = \bm{g}(\vd) - \bm{K}_{\!j}\vd$ the joint force correction. No inverse of $\mM$ is formed anywhere.

The panel is unloaded, so the equilibrium is $\vw_0 = \bm{0}$ and $\vs(\bm{0}) = \bm{0}$, and the residual is expanded about it,
\begin{equation}
  \bm{R}(\vw) \,\approx\, \bm{A}\,\Delta\vw \,+\,
  \tfrac{1}{2}\,\bm{B}_2\!\left(\Delta\vw,\Delta\vw\right) \,+\,
  \tfrac{1}{6}\,\bm{C}_3\!\left(\Delta\vw,\Delta\vw,\Delta\vw\right) \,+\,
  \mathcal{O}\!\left(|\Delta\vw|^{4}\right),
  \label{eq:taylor}
\end{equation}
where $\bm{B}_2$ and $\bm{C}_3$ are the second and third Jacobian operators, symmetric multilinear functions of their arguments. The system is projected onto $m$ eigenvectors of the pencil $\bm{A}\bm{\Phi}_i = \lambda_i \bm{E}\bm{\Phi}_i$ through $\Delta\vw = \bm{\Phi}\bm{z} + \bar{\bm{\Phi}}\bar{\bm{z}}$ and premultiplied by the left eigenvectors normalised so that $\bm{\Psi}_i^{\mathsf{T}}\bm{E}\bm{\Phi}_j = \delta_{ij}$, which gives $m$ complex ordinary differential equations
\begin{equation}
  \dot{z}_i \,=\, \lambda_i z_i \,+\, \bm{\Psi}_i^{\mathsf{T}}\vf(t)
  \,+\, \bm{\Psi}_i^{\mathsf{T}}\!\left[
  \tfrac{1}{2}\,\bm{B}_2(\Delta\vw,\Delta\vw) \,+\,
  \tfrac{1}{6}\,\bm{C}_3(\Delta\vw,\Delta\vw,\Delta\vw)\right].
  \label{eq:romdyn}
\end{equation}
For classical damping this complex state-space form collapses exactly onto an equivalent real second-order form in the modal amplitudes, which is the form actually integrated. The coefficients are computed once, after the equilibrium, the eigenvalues and the eigenvectors are known.

\subsection{Matrix-free coefficients and the step size}
\label{sec:nromeps}

The higher-order operators are never assembled. Only their action on the basis vectors is needed, and that is obtained matrix-free by central differences of the full-order residual along eigenvector directions \citep{woodgate2007matrixfree},
\begin{equation}
  \bm{B}_2(\bm{x},\bm{x}) = \frac{\bm{R}_1 - 2\bm{R}_0 + \bm{R}_{-1}}{\epsilon^{2}}, \qquad
  \bm{C}_3(\bm{x},\bm{x},\bm{x}) = \frac{-\bm{R}_3 + 8\bm{R}_2 - 13\bm{R}_1 + 13\bm{R}_{-1} - 8\bm{R}_{-2} + \bm{R}_{-3}}{8\,\epsilon^{3}},
  \label{eq:mfree}
\end{equation}
with $\bm{R}_l = \bm{R}(\vw_0 + l\epsilon\,\bm{x})$. Mixed arguments follow from polarization identities and complex arguments by splitting into real and imaginary parts and using multilinearity.

\begin{figure}[htbp]
  \centering
  \includegraphics[width=0.55\linewidth]{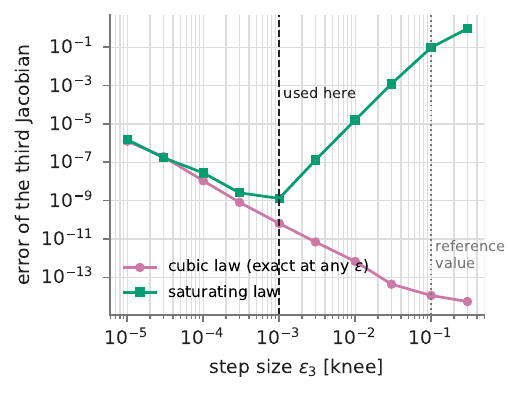}
  \caption{Step-size convergence of the matrix-free third Jacobian, measured against the analytic tensor. The cubic law is differentiated exactly at every step size, which isolates the round-off branch. The saturating law shows the expected fourth-order truncation branch.}
  \label{fig:eps}
\end{figure}

These differences suffer from truncation error when $\epsilon$ is too large and from round-off when it is too small, and the third-order stencil, which divides by $\epsilon^3$, is the sensitive one. The directions used here are normalised to unit knee deformation, so $\epsilon_3$ is the fraction of the knee that the stencil samples and its value is comparable across models. \Cref{fig:eps} shows the two branches cleanly. For the cubic law the stencil is exact at every step size, since a six-point formula differentiates a cubic exactly, and only the round-off branch is visible, rising as $\epsilon_3^{-3}$. For the saturating law the truncation branch falls as $\epsilon_3^{4}$, the two cross at $\epsilon_3 = 10^{-3}$, and the best attainable error there is $1.3\times10^{-9}$. The value $\epsilon_3 = 10^{-1}$ inherited from the aeroelastic application is $9.6\%$ in error on this normalisation, so $\epsilon_3 = 10^{-3}$ is used throughout. This is a reusable observation, since the appropriate step size is a property of the normalisation and not of the code.

\subsection{What sets the size of the basis}
\label{sec:nrombasis}

\begin{figure}[htbp]
  \centering
  \includegraphics[width=\linewidth]{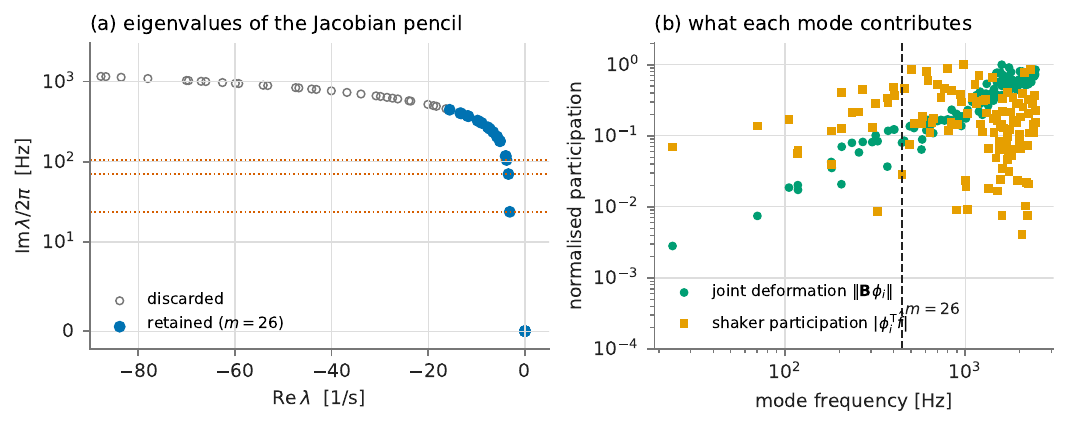}
  \caption{(a) Eigenvalues of the pencil with non-negative imaginary part, and the retained set. Dotted lines mark the three measured resonances. (b) Normalised participation of each mode in the shaker force and in the joint deformation. The joint deformation is carried by the high modes.}
  \label{fig:eigs}
\end{figure}

\Cref{fig:eigs} shows the eigenvalues of the pencil and what each mode contributes. The retained set is the lowest $m$, which for the elastic modes means those whose damped natural frequencies lie in and above the measured band. The second panel is the part that matters. The shaker participation is spread over the whole spectrum, while the joint deformation carried by a mode, $\|\mB\vphi_i\|$, increases with frequency. The joint deformation of this panel is not a low-frequency quantity.

\begin{figure}[htbp]
  \centering
  \includegraphics[width=\linewidth]{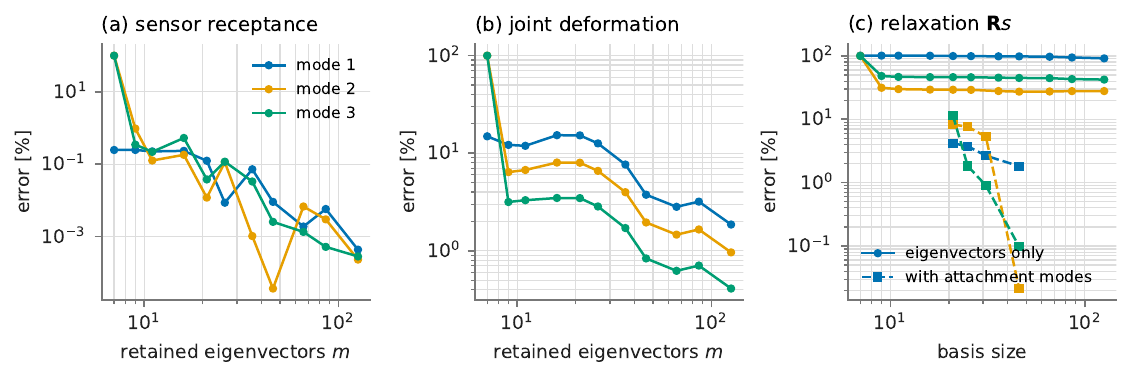}
  \caption{What sets the size of the basis. The sensor receptance and the joint deformation converge with the number of retained vectors, at very different rates, whereas the joint-to-joint receptance, the operator through which a joint force relaxes its own deformation, does not converge at any practical size, and it is the one the nonlinear model depends on.}
  \label{fig:basis}
\end{figure}

Three quantities have to be converged, and they converge at very different rates, \Cref{fig:basis}. The first is the sensor receptance. Measured against a direct sparse solve of the full model and taking the worst of the fifteen outputs, eleven vectors, that is the six rigid-body modes and the five elastic modes below $120~\mathrm{Hz}$, give $1.2\%$ at $23.8~\mathrm{Hz}$ and $3.8\%$ at $71.4~\mathrm{Hz}$, forty-six give $0.17\%$ and $0.51\%$, and $126$ give $0.024\%$ and $0.070\%$, an order of magnitude below the mesh bias of the underlying model. Dropping the six rigid-body modes from a basis of twenty-six costs $23.4\%$ and $7.7\%$, so for this free--free panel they are not optional. The second is the joint deformation produced by the shaker, which is the argument of the nonlinear law, and it converges an order of magnitude more slowly, reaching $1.9\%$ at $126$ vectors. The third does not converge at all.

That third quantity is the joint-to-joint receptance $\bm{R}(\omega) = \mB\bm{Z}(\omega)^{-1}\mB^{\mathsf{T}}$. It is the operator through which a joint's own nonlinear force relaxes its deformation, so it enters the reduced model wherever the nonlinearity does. Evaluated at the three resonances, a basis of $126$ eigenvectors reproduces on average $2.1$, $2.2$ and $2.4\%$ of its diagonal, and the error in $\bm{R}\vs$ for the force pattern actually present at resonance stalls at $91$, $28$ and $42\%$. Adding more modes does not help, because the missing content is the local static flexibility of the fastener region and no global mode contains it. This is the same observation that made the residual correction of \Cref{eq:rayleigh} non-optional, seen now from the side of the basis.

The consequence is severe. With that relaxation missing, the reduced model is not merely inaccurate, it is unphysical, overpredicting the hardening shift of the fundamental by a factor of thirty and returning a value that exceeds the rigid-joint limit of the panel. It is worth being explicit that the linear check passes comfortably at the same basis size, so sizing a basis on a linear convergence study is exactly what produces the wrong answer here. The linear response never loads a joint by its own force, so it never tests the operator the nonlinear model depends on.

Enriching the basis with residual attachment modes for the nonlinear force directions recovers the missing flexibility systematically and brings the reduced model back to the full-order answer. Five elastic modes generate $35$ force directions, of which ten carry $99.74\%$ of the energy, and the error in $\bm{R}\vs$ falls from $91$, $28$ and $42\%$ with eigenvectors alone to $1.8$, $0.02$ and $0.10\%$ with eleven modes and thirty-five attachment vectors. The cost of that repair is a basis three to four times larger, and by the time the model is accurate above $10~\mathrm{N}$ its per-frequency cost has fallen back to that of the condensed full-order solution. The reduction is fastest where it is least accurate, which is the difficulty the next section is designed to avoid.

\subsection{Keeping the residual flexibility out of the basis}
\label{sec:nromlocal}

The enrichment works, but it works by making the basis larger, which is what the reduction was trying to avoid. There is a second route, and on this structure it is the better one. Instead of adding the missing flexibility to the basis, keep it outside.

Split the response into the retained modal part and a static remainder, $\vu = \bm{\Phi}\vq + \bm{\mathcal{R}}(\vf_{\mathrm{ext}} + \mB^{\mathsf{T}}\vf_{\mathrm{nl}})$ with $\vf_{\mathrm{nl}} = -\vs(\vd)$. The joint deformation then satisfies
\begin{equation}
  \vd \,+\, \mRres\,\vs(\vd) \,=\, \mB\bm{\Phi}\,\vq \,+\, \bm{e}_{\mathrm{res}}F_0,
  \qquad
  \mRres = \mB\bm{\mathcal{R}}\mB^{\mathsf{T}},
  \label{eq:localcond}
\end{equation}
where $\mRres$ is the part of the joint receptance the retained modes do not carry. Written this way the missing flexibility never enters the basis at all, it enters a constraint that ties $\vd$ to the modal amplitudes.

What makes this cheap is a property of the operator rather than of the formulation. $\mRres$ is the local static flexibility of the fastener region, and on this panel it is almost perfectly local, its $2\times2$ diagonal blocks, one per joint element, carrying $99.7\%$ of its Frobenius norm. Dropping the remainder makes \Cref{eq:localcond} a set of $156$ independent two-variable problems, solved elementwise by a vectorised Newton that converges in two or three iterations, with a block-diagonal tangent
\begin{equation}
  \frac{\partial\vd}{\partial\vq} \,=\,
  \left(\mI + \mRres\mSp\right)^{-1}\mB\bm{\Phi},
  \qquad \mSp = \partial\vs/\partial\vd,
  \label{eq:localtangent}
\end{equation}
inverted in closed form two by two. The dynamic state stays the $m$ modal amplitudes, the joints are statically condensed onto them with their own flexibility, and the basis never has to grow. Building $\mRres$ costs one shifted factorisation and $312$ static solves, $2.7~\mathrm{s}$ in total, and it is frequency independent, so it is built once.

The effect on the quantity that governs everything is immediate. With eleven vectors the error in the joint receptance, measured against the exact $\mB\bm{Z}^{-1}\mB^{\mathsf{T}}$, falls from $99.96$, $99.86$ and $98.79\%$ at the three resonances to $0.085$, $0.146$ and $0.289\%$. Those are the numbers that thirty-five attachment vectors were needed to approach, obtained here with a basis a quarter of the size and one extra factorisation. Because the condensation evaluates the constitutive law itself rather than a truncated expansion of it, it is also not restricted to laws that admit a Taylor series, and the saturating and bilinear laws, for which no third-order model exists past the knee, are reachable.

\subsection{How accurately the discarded operator has to be represented}
\label{sec:dial}

Discarding the off-block part of $\mRres$ is an approximation, and it is not uniformly harmless. Keeping the diagonal blocks alone drops $8.3\%$ of $\|\mRres\|$ in Frobenius norm, the retained and discarded parts adding in quadrature, so that the $99.7\%$ carried by the blocks and the $8\%$ left outside them are the same statement. That same discarded norm costs $0.06\%$ in the joint deformation while the joints stay below their slip knee and $39\%$ once they slip together.

\begin{figure}[htbp]
  \centering
  \includegraphics[width=\linewidth]{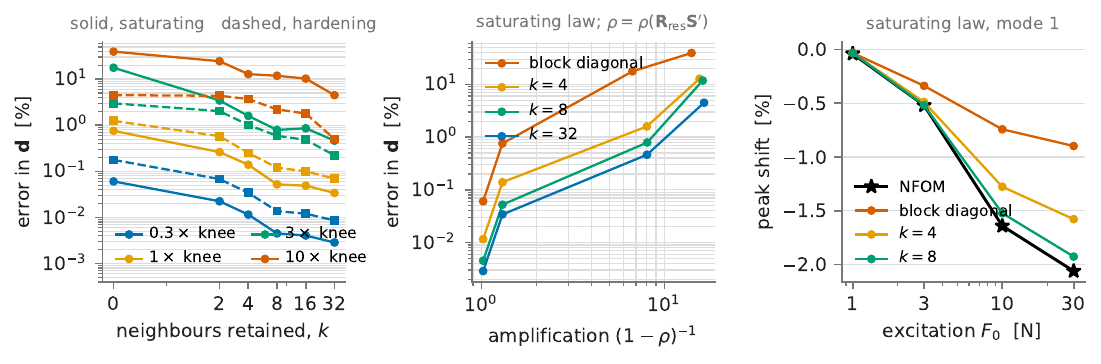}
  \caption{What a $k$-nearest-neighbour condensation of $\mRres$ recovers, against a dense Newton on the complete $\mRres$. Left, error in the joint deformation against the number of neighbouring joint elements retained, at four amplitudes, for the saturating law (solid) and the hardening law (dashed). Centre, the same errors against the amplification $(1-\rho)^{-1}$ at the solution. Right, what the retained pattern does to the observable, the peak shift of the fundamental under the saturating law.}
  \label{fig:neighbours}
\end{figure}

What the error follows is therefore not the size of what is dropped but the conditioning of \Cref{eq:localcond}. The spectral radius $\rho = \rho(\mRres\mSp)$ at the solution grows from $0.02$ to $0.94$ over the same range, and any error in the joint receptance is amplified at the solution by $(1-\rho)^{-1}$. \Cref{fig:neighbours} is that statement measured, against a dense Newton on the complete $\mRres$. A truncation of $\mRres$ is safe exactly while the joints stay near their knee, and no truncation is safe once they slip together, which is precisely the regime the nonlinear study is about.

The remedy is not a better preconditioner. A block-preconditioned iteration on the full coupling converges only while $\rho < 1$, and $\rho$ is not small in the regime of interest. Measured at the solution of every case of this study it reaches $0.92$ to $0.95$ for the two softening laws, an amplification of twelve to twenty, and for the hardening law it exceeds one everywhere, from $1.3$ on the fundamental at $1~\mathrm{N}$ to $66$ on mode 2 at $30~\mathrm{N}$, so such an iteration diverges for every hardening case here.

What works is to solve a retained pattern exactly. Keeping, for each joint element, the $k$ nearest other elements of $\mRres$ and reordering the retained pattern by reverse Cuthill--McKee makes it banded, with eight neighbours giving a half-bandwidth of $35$ on the $312$ joint unknowns, so the Newton system of \Cref{eq:localtangent} is factored in banded storage at a cost of order $n_{\mathrm{nl}}b^2$ rather than $n_{\mathrm{nl}}^3$, which is $1.5$ against $20$ megaflop per time sample, a factor of thirteen once the fill from partial pivoting is counted. The time samples of the harmonic balance are independent, so they are stacked into one block-diagonal band and factored together, and the band is assembled directly from shifted windows of the retained matrix rather than by forming and stripping a dense Jacobian. Being an exact Newton, it does not care about $\rho$, converging in at most ten iterations to a constraint residual of $4\times10^{-14}$, with a tangent correct to central-difference accuracy, over the three laws and up to $\rho = 14.2$.

\subsection{Results and cost}
\label{sec:nromresults}

\begin{figure}[htbp]
  \centering
  \includegraphics[width=\linewidth]{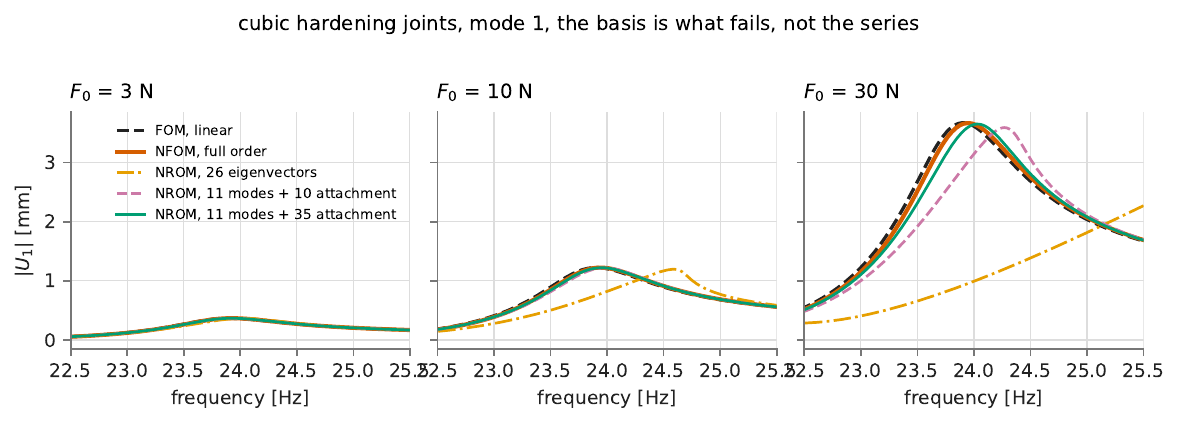}
  \caption{Cubic hardening joints at $30~\mathrm{N}$, the linear full-order model, the nonlinear full-order model, the linear reduced model and the third-order nonlinear reduced model, at the sensor with the largest response in each band.}
  \label{fig:frf}
\end{figure}

\Cref{tab:nromlocal} reports both configurations against the nonlinear full-order model. The hardening law, which the block-diagonal condensation already reproduced to $0.21$ percentage points, is reproduced to $0.01$ on every mode and at every level once the coupling is restored. On the fundamental the two softening laws, whose fully slipped shift the block-diagonal model underpredicted by a factor of two and a third, are recovered to $0.13$ percentage points. \Cref{fig:frf} shows the comparison of the models around the fundamental.

\Cref{tab:cost} gives the cost, timed as a parametric study would pay it, with everything built once listed separately. The anchor is a direct sparse solve of the full model at $1000~\mathrm{ms}$ per frequency. The condensed harmonic balance is already a reduction of that at $306~\mathrm{ms}$, though an expensive one to set up. Against those, the linear reduced model costs $1.6~\mathrm{ms}$ and the third-order model $46~\mathrm{ms}$ on a basis of twenty-odd vectors, but that basis is the one \Cref{sec:nrombasis} shows to be unusable, and the enriched basis that is accurate costs $399~\mathrm{ms}$, no better than the model it replaces.

The local model does not have one cost, because the neighbour count sets it. On eleven vectors the block-diagonal condensation costs $118~\mathrm{ms}$ per frequency, a factor of $2.6$ below the condensed full-order solution and $8.5$ below a direct one, four neighbours cost $288~\mathrm{ms}$, eight cost $595~\mathrm{ms}$ and sixteen cost $1109~\mathrm{ms}$. Read together with \Cref{tab:nromlocal} this is an exchange and not a free improvement. The configuration that reproduces the fully slipped shift of the fundamental to a tenth of a percentage point costs about twice the model it is reducing, and the configuration that is cheaper than that model is the one which underpredicts the same shift by a factor of $2.3$. Four neighbours sit near the crossing point, recovering about two thirds of the accuracy at parity of cost. The neighbour count is best read as an accuracy dial rather than as a speed-up, and the measured figures understate the formulation, since the arithmetic is far smaller than the full-order factorisation and it is the overhead of many small array operations, not the method, that limits the measured gain.

One limit should be stated plainly, because separating it from the coupling is what the neighbour study was for. Raising $k$ converges the condensation onto its own full-coupling limit in every case, and what differs between modes is how far that limit sits from the full-order answer. On the fundamental at $30~\mathrm{N}$ the sequence $-0.90$, $-1.58$, $-1.93$, $-2.10$ for $k = 0$, $4$, $8$, $32$ brackets the full-order $-2.06$ and settles within $0.04$ percentage points of it. On mode 2 the same sequence runs $-1.61$, $-3.59$, $-5.51$ past a full-order $-4.87$ and settles beyond it. That difference is not the conditioning, since $\rho$ is $0.94$ in both, and it is not the reference frequency at which the residual is built, since moving that from $70.3$ to $60~\mathrm{Hz}$ changes the result by $0.03$ percentage points. What is left is the eleven-vector representation of the joint receptance itself, which \Cref{sec:nromlocal} reports as $0.085\%$ at the fundamental and $0.146\%$ at mode 2, amplified by the same $(1-\rho)^{-1}$. The off-block coupling is therefore no longer the limiting approximation of this formulation, and the basis is.

\begin{table}[htbp]
  \centering
  \caption{Peak-frequency shift on eleven vectors against the nonlinear full-order model, in the two configurations of the local condensation, at $30~\mathrm{N}$. $\rho$ is the spectral radius at the solution of that case. The full-order reference here is recomputed on the finer frequency grid of the reduction study, which accounts for the differences of $0.01$ percentage points against \Cref{tab:nlsummary}.}
  \label{tab:nromlocal}
  \small
  \begin{tabular}{l c r r r r r}
    \toprule
    & & & & \multicolumn{1}{c}{block diagonal} & \multicolumn{2}{c}{eight neighbours} \\
    \cmidrule(lr){5-5}\cmidrule(lr){6-7}
    law & mode & $\rho$ & NFOM [\%] & $\Delta f$ [\%] & $\Delta f$ [\%] & error [pp] \\
    \midrule
    cubic hardening   & 1 & 25   & $+0.18$ & $+0.20$ & $+0.18$ & $+0.00$ \\
                      & 2 & 66   & $+0.48$ & $+0.61$ & $+0.49$ & $+0.01$ \\
                      & 3 & 59   & $+0.91$ & $+1.12$ & $+0.92$ & $+0.01$ \\
    \midrule
    smooth saturating & 1 & 0.94 & $-2.06$ & $-0.90$ & $-1.93$ & $+0.13$ \\
                      & 2 & 0.95 & $-4.87$ & $-1.61$ & $-3.59$ & $+1.28$ \\
                      & 3 & 0.95 & $-9.17$ & $-3.57$ & $-8.38$ & $+0.79$ \\
    \bottomrule
  \end{tabular}
\end{table}

\begin{table}[htbp]
  \centering
  \caption{Cost of the full-order and reduced-order solutions on the same machine. The per-frequency figures are the dynamic response alone, everything built once being listed separately, since a parametric study pays it once.}
  \label{tab:cost}
  \small
  \begin{tabular}{l l r}
    \toprule
    stage & item & cost \\
    \midrule
    \multirow{4}{*}{offline}
      & full-order condensed model, modal receptance and reference solves & $520$ s \\
      & reduced model, eigensolve of $126$ modes & $7$ s \\
      & reduced model, third-order coefficients, $26$ vectors & $5$ s \\
      & reduced model, residual flexibility, one factorisation and $312$ solves & $2.7$ s \\
    \midrule
    \multirow{7}{*}{per frequency}
      & FOM linear, direct sparse solve of the $18{,}804$ degrees of freedom & $1000$ ms \\
      & NFOM, condensed harmonic balance on the $312$ joint unknowns & $306$ ms \\
      & ROM linear, $26$ eigenvectors & $1.6$ ms \\
      & NROM, $26$ eigenvectors & $46$ ms \\
      & NROM, $11$ modes $+\,35$ attachment vectors & $399$ ms \\
      & NROM, local residual flexibility, block diagonal & $118$ ms \\
      & the same, $8$ neighbours retained (bandwidth $35$) & $595$ ms \\
    \bottomrule
  \end{tabular}
\end{table}

\subsection{Verification of the reduction}
\label{sec:nromverify}

\Cref{tab:nromverify} lists the verification of the reduction, at the level of the basis, of the Taylor coefficients, of the reduced solver and of the condensation. Three entries carry most of the weight. The matrix-free third Jacobian is checked against an analytically assembled tensor for a law where that tensor exists, which tests \Cref{eq:mfree} and the polarization identities together. The Taylor model is checked against an exact-law modal Galerkin model on the same basis, which separates the error of the expansion from the error of the basis, and for a cubic law the two agree to $2.7\times10^{-11}$ because the series is then exact. And the banded assembly of \Cref{sec:dial} is checked against the dense Jacobian it replaces, which it reproduces to machine precision, so that the neighbour condensation is an exact solve of a truncated operator rather than an approximate solve of the full one.

\begin{table}[htbp]
  \centering
  \caption{Verification of the model order reduction. Every entry is a relative deviation from an independent reference, an analytic tensor, an analytic receptance, an independently coded evaluation of the same quantity, or a property that holds exactly by construction. The condensation entries are the worst case over the three smooth laws, over neighbour counts $4$, $8$ and $16$, and over joint deformations reaching a hundred times the slip knee.}
  \label{tab:nromverify}
  \small
  \begin{tabular}{l p{0.50\linewidth} r}
    \toprule
    level & check & deviation \\
    \midrule
    \multirow{3}{*}{basis}
      & quadratic eigenvalue residual of the retained pairs & $6.2\times10^{-9}$ \\
      & bi-orthonormality of the left and right eigenvectors & $1.7\times10^{-11}$ \\
      & rigid-body modes carry no joint deformation & $1.5\times10^{-11}$ \\
    \midrule
    \multirow{3}{*}{coefficients}
      & second-order tensor vanishes by isotropy, cubic law & $0$ exactly \\
      & matrix-free third Jacobian vs the analytic tensor & $1.1\times10^{-14}$ \\
      & third-order model vs the exact cubic law at the knee & $1.1\times10^{-14}$ \\
    \midrule
    \multirow{3}{*}{reduced solver}
      & linear reduced harmonic balance vs its analytic receptance & $2.8\times10^{-16}$ \\
      & Taylor model vs the exact-law model, cubic, $30~\mathrm{N}$ & $2.7\times10^{-11}$ \\
      & state-space form vs the equivalent real second-order form & $5.1\times10^{-15}$ \\
    \midrule
    \multirow{3}{*}{condensation}
      & banded assembly vs the dense Jacobian it replaces & $1.7\times10^{-15}$ \\
      & constraint \eqref{eq:localcond} at the converged deformation & $4.2\times10^{-14}$ \\
      & analytic tangent of the condensation vs central differences & $1.2\times10^{-8}$ \\
    \bottomrule
  \end{tabular}
\end{table}

\subsection{Where the expansion is valid}
\label{sec:nromvalid}

A second limitation belongs to the expansion rather than to the basis. A Taylor series about the undeformed state describes a joint law only while the joints stay near their knee, and the levels at which joint nonlinearity is interesting in this panel are one to two orders of magnitude beyond it. A cubic law is reproduced exactly at any amplitude because it is itself a cubic. A saturating law is not, since the Taylor series of $\tanh$ has a finite radius of convergence and truncating at third order produces a softening cubic that is unbounded below, so the reduced model has no bounded periodic solution once the joints pass the knee. The bilinear law has no Taylor expansion at the origin because it is not differentiable at the knee, and the friction law is outside the framework entirely because it carries an internal state.

Of the four laws, only the cubic one lies inside the assumptions of the Taylor model over the whole range of this study, which is why the comparison of \Cref{sec:nromresults} is made on it. This restriction belongs to the expansion and not to the reduction. The local condensation of \Cref{sec:nromlocal} evaluates the law itself and reaches the saturating and bilinear cases as well. Only the Jenkins law remains outside both, because a hysteretic element carries an internal state that neither formulation condenses, and any reduced model intended for the friction case will have to keep the nonlinearity in a form that saturates, which points towards the exact condensation of \Cref{sec:nlsolution} or towards a basis adapted to the operating point rather than fixed, as in \citep{hollins2026projection}.

\section{Conclusions}
\label{sec:conclusions}

An open shell finite-element model of a stiffened aluminium wingbox panel has been built from mixed-interpolation flat-shell elements with the plate-to-stiffener joints at the $78$ physical fastener positions, and calibrated against measured modal data by a six-parameter sensitivity update constrained to $\pm20\%$, reaching $1.1$ to $4.3\%$ in frequency with modal-assurance values of $0.88$ or better. Folded-shell benchmarks proved to be the decisive verification tests for this class of model, since a flat-plate check cannot detect an error in the rotation transformation across a fold line, and they are recommended for any in-house shell code applied to stiffened panels.

Making the fasteners of that model nonlinear shows that joint nonlinearity enters the response strongly asymmetrically. Hardening is almost invisible, whereas softening and slip move the first three resonances by $-2.1$, $-4.9$ and $-9.2\%$, and friction removes up to $80\%$ of the resonant peak before it recovers. Two independent solutions, one in the time domain on every degree of freedom and one in the frequency domain on the joints alone, agree to $5.5\times10^{-4}$, so the effect is a property of the model rather than of a solver, and the shifts are bounded by a fully saturated limit they approach to within $0.3\%$ while remaining proportional to a joint strain-energy fraction that never exceeds $2\%$.

Applying a projection-based nonlinear reduction to the same model locates a criterion that a jointed structure imposes on any reduced basis. The binding quantity is neither the linear response, which eleven vectors reproduce to better than $0.01\%$, nor the joint deformation the shaker produces, but the receptance of the structure between the joints, of which $126$ global eigenvectors carry about $2\%$, and without which the reduced model is unphysical. Condensing that operator instead of spanning it avoids the collision between accuracy and cost that basis enrichment runs into, and how accurately it must be condensed is set by one quantity, since any error in the joint receptance is amplified at the solution by $(1-\rho)^{-1}$. The same eight percent of the operator costs $0.06\%$ below the slip knee and $39\%$ once the joints slip together. Retaining the eight nearest joint elements and solving that band as an exact Newton, exact rather than preconditioned since $\rho$ exceeds unity for every hardening case here, reproduces the full-order shift to $0.01$ percentage points for hardening and $0.13$ for a fully slipped saturating law, at a cost of about twice the model it reduces, so the neighbour count is an accuracy dial rather than a speed-up.

The result is not specific to this panel. When a nonlinearity is localised, the operator a modal basis cannot represent is localised too, and it is cheaper to condense it than to span it, provided the accuracy of that condensation is judged by how far the joints slip rather than by how small the discarded operator looks.

\section*{Reproducibility}
All results are produced by a pure-Python implementation released with this paper, organised as the shell element and updating machinery, the nonlinear full-order solvers, and the reduced-order models, each with a README giving the exact sequence of commands and with verification scripts that regenerate every figure and table.

\bibliographystyle{unsrtnat}
\bibliography{references}

\end{document}